\documentclass[aps,prb,reprint,nolongbibliography,noeprint,superscriptaddress]{revtex4-2}
\usepackage{amsfonts,amssymb,amsmath,bm,graphicx,color}
\usepackage[colorlinks=true,citecolor=blue,linkcolor=blue,urlcolor=blue]{hyperref}
\allowdisplaybreaks[4]
\pdfmapline{=cmr12 CMR12 <cmr12.pfb <f7b6d320.enc}
\newcommand\db[1]{\dot {\bm #1}}
\newcommand\hb[1]{\hat {\bm #1}}
\newcommand\hc[1]{\hat {\cal #1}}
\newcommand\pphantom[1]{\protect\phantom{#1}}

\DeclareMathOperator\sgn{sgn}
\begin{document}
%--- Front Matter
\title{Chiral vortical effect in relativistic and nonrelativistic systems}
\author{Atsuo Shitade}
\affiliation{Institute for Molecular Science, Aichi 444-8585, Japan}
\author{Kazuya Mameda}
\affiliation{Theoretical Research Division, Nishina Center, RIKEN, Wako, Saitama 351-0198, Japan}
\author{Tomoya Hayata}
\affiliation{Department of Physics, Keio University, 4-1-1 Hiyoshi, Kohoku-ku, Yokohama 223-8521, Japan}
\date{\today}
\begin{abstract}
  We formulate the chiral vortical effect (CVE) and its generalization called generalized vortical effect using the semiclassical theory of wave packet dynamics.
  We take the spin-vorticity coupling into account and calculate the transport charge current by subtracting the magnetization one from the Noether local one.
  We find that the transport charge current in the CVE always vanishes in relativistic chiral fermions.
  This result implies that it cannot be observed in transport experiments
  in condensed matter systems such as Dirac/Weyl semimetals with the pseudo-Lorentz symmetry.
  We also demonstrate that the anisotropic CVE can be observed in nonrelativistic systems
  that belong to the point groups $D_n, C_n (n = 2, 3, 4, 6)$, and $C_1$, such as $n$-type tellurium.
\end{abstract}
\maketitle
%--- Main Matter
\section{Introduction} \label{sec:introduction}
Quantum anomalies play important roles in high energy and condensed matter physics.
In a relativistic system of chiral fermions, the chiral symmetry in the classical action breaks down
when the theory is quantized in the presence of electromagnetic fields,
which is known as a chiral anomaly~\cite{PhysRev.177.2426,Bell1969}.
Such a system is realized in quark-gluon plasmas in heavy-ion collision experiments~\cite{ARSENE20051,BACK200528,ADAMS2005102,ADCOX2005184}.
A famous consequence of the chiral anomaly is the chiral magnetic effect (CME)~\cite{PhysRevD.22.3080,KHARZEEV2006260,KHARZEEV200767,KHARZEEV2008227,PhysRevD.78.074033};
the charge current ${\bm j}$ flows parallel to a magnetic field ${\bm B}$ as ${\bm j} \propto \mu_5 {\bm B}$ with the chiral chemical potential $\mu_5 = (\mu_{\rm L} - \mu_{\rm R})/2$,
where $\mu_{\rm L/R}$ are the chemical potentials for the left/right-handed fermions, respectively.
However, in equilibrium, the charge current is forbidden by the Bloch-Bohm theorem~\cite{PhysRev.75.502,JPSJ.65.3254,PhysRevD.92.085011}.
In fact, nonzero $\mu_5$ cannot be realized in equilibrium.
{Under an applied electric field} ${\bm E}$ as well as a magnetic field, the chiral imbalance proportional to ${\bm E} \cdot {\bm B}$ is generated out of equilibrium
and causes the negative magnetoresistance (NMR) via the CME~\cite{NIELSEN1983389,PhysRevB.88.104412}.

In condensed matter physics, the chiral anomaly has been studied in Dirac/Weyl semimetals~\cite{RevModPhys.90.015001}.
They are three-dimensional materials with gapless electronic excitations protected by topology and symmetry.
According to the Nielsen-Ninomiya theorem~\cite{NIELSEN1983389}, such nodes always appear in pairs with the opposite chiralities in lattice systems,
and hence the CME does not occur in equilibrium.
A simple proof using the periodicity of the Brillouin zone was also given in Ref.~\cite{PhysRevLett.111.027201}.
Instead, the Fermi arc surface states~\cite{PhysRevx.5.031013,xu613,Lv2015,Yang2015a,*Yang2015b,xu2015,xue1501092,PhysRevLett.116.096801,xu2016}
and the aforementioned NMR~\cite{PhysRevx.5.031023,Du2016,Zhang2016,PhysRevB.93.121112} were observed in TaAs and its family 
soon after the first-principles predictions~\cite{PhysRevx.5.011029,Huang2015},
although it was pointed out that the observed NMR may be attributed to the current jetting effect~\cite{Arnold2016,dosReis2016}.

The chiral vortical effect (CVE) is also an anomaly-related transport phenomenon~\cite{PhysRevD.20.1807,PhysRevD.21.2260,KHARZEEV200767,Erdmenger2009,Banerjee2011,PhysRevLett.103.191601,PhysRevLett.107.021601,Landsteiner2011,PhysRevLett.113.182302};
the charge current flows parallel to the vorticity ${\bm \omega}$ as ${\bm j} \propto \mu \mu_5 {\bm \omega}$ with the chemical potential $\mu = (\mu_{\rm L} + \mu_{\rm R})/2$.
Although the conductivity is proportional to $\mu_5$, the CVE has been believed to occur in equilibrium in noncentrosymmetric Weyl semimetals~\cite{PhysRevB.89.035142,PhysRevB.89.075124}.
Indeed, the Bloch-Bohm theorem is inapplicable to the CVE in equilibrium;
the theorem requires the thermodynamic limit, while the system with vorticity must be finite-sized so that the causality constraint is respected~\cite{PhysRevD.21.2260,PhysRevD.92.085011}.
Recently in the condensed matter context, an extension called generalized vortical effect (GVE) has been proposed~\cite{PhysRevResearch.2.032021},
in which the charge current is induced by a velocity gradient $\partial_{x^j} v_l$ as $j^i = q \sigma_{\rm v}^{ijl} \partial_{x^j} v_l/2$,
for some noncentrosymmetric materials in the hydrodynamic regime such as a Weyl semimetal WP$_2$~\cite{Gooth2018}.
The GVE includes the CVE as the completely antisymmetric case and the anisotropic CVE discussed later.

In this paper, we formulate the CVE and GVE using the semiclassical theory of wave packet dynamics~\cite{RevModPhys.82.1959}.
This formalism correctly describes the anomalous Hall effect~\cite{PhysRevLett.75.1348,PhysRevB.53.7010},
the orbital magnetization~\cite{PhysRevLett.95.137204,*PhysRevLett.95.169903,PhysRevLett.97.026603}, the anomalous Nernst effect~\cite{PhysRevLett.97.026603}, and many others.
One advantage of using the theory is to separate the effects of the spin-vorticity coupling (SVC), magnetic moment, and Berry curvature at the expense of the explicit Lorentz covariance,
which makes the physical origins of the CVE and GVE evident.
We take the SVC into account and calculate the transport charge current by subtracting the magnetization one from the local one~\cite{PhysRevResearch.2.032021,PhysRevLett.97.026603,PhysRevB.55.2344}.
Regarding the relativistic system of chiral fermions,
we find that the transport charge current always vanishes whether the system is in equilibrium or not because the contributions of the SVC and Berry curvature cancel each other.
In other words, the CVE cannot be observed in transport experiments.
On the other hand, we show that the anisotropic CVE can be observed in nonrelativistic systems that belong to the point groups $D_n, C_n (n = 2, 3, 4, 6)$, and $C_1$.
We estimate the induced charge current using the low-energy effective Hamiltonian of $n$-type tellurium.

\section{Semiclassical theory} \label{sec:semiclassical}
We consider a moving fluid observed in an inertial frame using the semiclassical theory of wave packet dynamics.
The Hamiltonian of the system is ${\hc H}({\bm p})$,
whose eigenvalues and eigenstates are $\epsilon_n({\bm p})$ and $| u_n({\bm p}) \rangle$.
The equations of motion are
\begin{subequations} \begin{align}
  {\db x}
  = & {\bm \nabla}_p \epsilon_n({\bm p}), \label{eq:semiclassical1a} \\
  {\db p}
  = & 0. \label{eq:semiclassical1b}
\end{align} \label{eq:semiclassical1}\end{subequations}
We assume that the distribution function $f$ is a function of
${\tilde \epsilon}_n({\bm p}, {\bm x}) = \epsilon_n({\bm p}) - {\bm v}({\bm x}) \cdot {\bm p} - {\bm \omega}({\bm x}) \cdot {\bm s}_n({\bm p})$
in the presence of the fluid velocity ${\bm v}({\bm x})$ and the vorticity ${\bm \omega}({\bm x}) = {\bm \nabla}_x \times {\bm v}({\bm x})/2$.
The last term corresponds to the SVC, and ${\bm s}_n({\bm p}) = \langle u_n({\bm p}) | {\hb s} | u_n({\bm p}) \rangle$ with ${\hb s}$ being the spin operator.
This distribution function may or may not be the solution of the collision-free Boltzmann equation,
\begin{equation}
  \partial_t f + {\db x} \cdot {\bm \nabla}_x f + {\db p} \cdot {\bm \nabla}_p f
  = 0. \label{eq:semiclassical2}
\end{equation}
In the latter case, we need to add the collision integral to the right-hand side of Eq.~\eqref{eq:semiclassical2}.
Nonetheless, we neglect the collision effect and focus on the intrinsic mechanisms of the CVE and GVE.

Hereafter we compute the charge currents.
We define the magnetic moment and Berry curvature as
\begin{subequations} \begin{align}
  {\bm m}_n({\bm p})
  = & -i \hbar^2 \langle {\bm \nabla}_p u_n({\bm p}) | [\epsilon_n({\bm p}) - {\hc H}({\bm p})] | {\bm \nabla}_p u_n({\bm p}) \rangle/2, \label{eq:semiclassical3a} \\
  {\bm \Omega}_n({\bm p})
  = & i \hbar^2 \langle {\bm \nabla}_p u_n({\bm p}) | \times | {\bm \nabla}_p u_n({\bm p}) \rangle. \label{eq:semiclassical3b}
\end{align} \label{eq:semiclassical4}\end{subequations}
The local charge current is then defined as~\cite{RevModPhys.82.1959}
\begin{align}
  {\bm j}_{\rm loc}({\bm x})
  = & q \sum_n \int \frac{d^3 p}{(2 \pi \hbar)^3} \{{\db x} f({\tilde \epsilon}_n({\bm p}, {\bm x})) \notag \\
  & + {\bm \nabla}_x \times [{\bm m}_n({\bm p}) f({\tilde \epsilon}_n({\bm p}, {\bm x}))/\hbar]\}. \label{eq:semiclassical4}
\end{align}
Since the magnetic moment ${\bm m}_n({\bm p})$ describes the self rotation of a wave packet,
it contributes to the local charge current when $f$ is nonuniform.
The second term was found in the chiral kinetic theory as well, which is indispensable for maintaining the Lorentz covariance~\cite{PhysRevLett.113.182302},
and the magnetic moment corresponds to the spin alignment of chiral fermions.
Indeed, when a wave packet is constructed from the positive energy bands of Dirac fermions, 
a nonAbelian extension of the magnetic moment is related to the expectation value of the spin operator~\cite{CHUU2010533}.

By substituting Eq.~\eqref{eq:semiclassical1a} into Eq.~\eqref{eq:semiclassical4} and expanding $f$ up to the first order with respect to ${\bm v}({\bm x})$ and ${\bm \omega}({\bm x})$,
we obtain
\begin{equation}
  j_{\rm loc}^i({\bm x})
  = q [n v^i({\bm x}) + {S^i}_j \omega^j({\bm x}) + \epsilon^{ijk} \partial_{x^j} M_{kl} v^l({\bm x})/2]. \label{eq:semiclassical5}
\end{equation}
The coefficients, one of which appears later, are
\begin{subequations} \begin{align}
  {S^i}_j
  = & \sum_n \int \frac{d^3 p}{(2 \pi \hbar)^3} \partial_{p_i} s_{n j}({\bm p}) f(\epsilon_n({\bm p})), \label{eq:semiclassical6a} \\
  M_{kl}
  = & -2 \sum_n \int \frac{d^3 p}{(2 \pi \hbar)^3} m_{n k}({\bm p}) p_l f^{\prime}(\epsilon_n({\bm p}))/\hbar, \label{eq:semiclassical6b} \\
  C_{kl}
  = & -2 \sum_n \int \frac{d^3 p}{(2 \pi \hbar)^3} \Omega_{n k}({\bm p}) p_l f(\epsilon_n({\bm p}))/\hbar, \label{eq:semiclassical6c}
\end{align} \label{eq:semiclassical6}\end{subequations}
which originate from the SVC, magnetic moment, and Berry curvature, respectively.

The local charge current itself cannot be measured in transport experiments.
Experimentally measured is the transport charge current
${\bm j}_{\rm tr}({\bm x}) = {\bm j}_{\rm loc}({\bm x}) - {\bm j}_{\rm mag}({\bm x})$~\cite{PhysRevResearch.2.032021,PhysRevLett.97.026603,PhysRevB.55.2344},
where ${\bm j}_{\rm mag}({\bm x}) = {\bm \nabla}_x \times {\bm M}({\bm x})$ is the magnetization charge current,
and ${\bm M}({\bm x})$ is the orbital magnetization.
The reason is as follows~\cite{PhysRevB.55.2344}:
Considering a cross section of the system, denoted by $S$, we readily find that the surface integral of ${\bm j}_{\rm mag}({\bm x})$ over $S$
turns into the vanishing line integral of ${\bm M}({\bm x})$ over $\partial S$ thanks to the Stokes theorem.
Thus, the net charge transport is determined by ${\bm j}_{\rm tr}({\bm x})$ alone.
In equilibrium, the orbital magnetization can be calculated as ${\bm M} = \partial P/\partial {\bm B}$~\cite{PhysRevLett.95.137204,*PhysRevLett.95.169903,PhysRevLett.97.026603},
where $P$ is the pressure.

When the fluid velocity and vorticity vary slowly, the orbital magnetization can be calculated as if they are uniform~\cite{PhysRevResearch.2.032021}, namely,
\begin{subequations} \begin{align}
  {\bm M}({\bm x})
  = & \frac{q}{\hbar} \sum_n \int \frac{d^3 p}{(2 \pi \hbar)^3} [{\bm m}_n({\bm p}) f({\tilde \epsilon}_n({\bm p}, {\bm x})) \notag \\
  & + {\bm \Omega}_n({\bm p}) f^{(-1)}({\tilde \epsilon}_n({\bm p}, {\bm x}))], \label{eq:semiclassical7a} \\
  f^{(-1)}(\epsilon)
  = & -\int_{\epsilon}^{\infty} d z f(z). \label{eq:semiclassical7b}
\end{align} \label{eq:semiclassical7}\end{subequations}
The second term of Eq.~\eqref{eq:semiclassical7a} originates from the Berry-phase correction of the density of states~\cite{PhysRevLett.95.137204,*PhysRevLett.95.169903,PhysRevLett.97.026603},
and $f^{(-1)}(\epsilon)$ satisfies $f^{(-1) \prime}(\epsilon) = f(\epsilon)$.
Hence, the mangetization charge current is
\begin{equation}
  j_{\rm mag}^i({\bm x})
  = q \epsilon^{ijk} \partial_{x^j} (M_{kl} + C_{kl}) v^l({\bm x})/2. \label{eq:semiclassical8}
\end{equation}
Together with Eq.~\eqref{eq:semiclassical5}, the transport charge current reads
\begin{align}
  j_{\rm tr}^i({\bm x})
  = q [n v^i({\bm x}) + {S^i}_j \omega^j({\bm x}) - \epsilon^{ijk} \partial_{x^j} C_{kl} v^l({\bm x})/2]. \label{eq:semiclassical9}
\end{align}
The first term is well known in hydrodynamics, and the third term was already obtained in Ref.~\cite{PhysRevResearch.2.032021}.
The second term was overlooked in condensed matter physics but turns out to be important below.
The generalized vortical conductivity, which is defined as $j_{\rm tr}^i = \sigma_{\rm v}^{ijl} \partial_{x^j} v_l/2$,
is $\sigma_{\rm v}^{ijl} = \epsilon^{kjl} S^i_{\pphantom{i} k} - \epsilon^{ijk} C_k^{\pphantom{k} l}$.
In the context of high energy physics, the chiral vortical conductivity is given by the scalar part of $S - C$.

\section{CVE for relativistic fermions} \label{sec:rel}
We apply our results to the relativistic system of chiral fermions.
The unperturbed Hamiltonian is ${\hc H}({\bm p}) = {\bm \sigma} \cdot {\bm p}$ with ${\bm \sigma}$ being the Pauli matrices for the spin degrees of freedom.
The eigenvalues are $\epsilon_{\sigma}({\bm p}) = \sigma p$ [$\sigma = \pm1$], and the corresponding eigenstates are
$| u_+({\bm p}) \rangle = [\cos \theta/2, e^{i \phi} \sin \theta/2]^{\rm T}$ and $| u_-({\bm p}) \rangle = [-e^{-i \phi} \sin \theta/2, \cos \theta/2]^{\rm T}$,
where $p$, $\theta$, and $\phi$ are the polar coordinates of ${\bm p}$.
We obtain ${\bm s}_{\sigma}({\bm p}) = \sigma (\hbar/2) {\bm e}_p,
{\bm m}_{\sigma}({\bm p}) = \hbar^2 {\bm e}_p/2 p$, and ${\bm \Omega}_{\sigma}({\bm k}) = -\sigma \hbar^2 {\bm e}_p/2 p^2$.
The distribution function of ${\tilde \epsilon}_\sigma({\bm p}, {\bm x})$ is the equilibrium one, which satisfies the collision-free Boltzmann equation~\eqref{eq:semiclassical2}.
The coefficients in Eq.~\eqref{eq:semiclassical6} are then evaluated as
${S^i}_j = (1/3) \sigma_{\rm v} \delta^i_{\pphantom{i} j}$, $M_{kl} = (2/3) \sigma_{\rm v} \delta_{kl}$, and $C_{kl} = (1/3) \sigma_{\rm v} \delta_{kl}$.
Here, $\sigma_{\rm v}$ is the chiral vortical conductivity computed with the local charge current,
i.e., $j_{\rm loc}^i=\sigma_{\rm v} \omega^i$, and reads
\begin{equation}
  \sigma_{\rm v}
  = \frac{1}{(2 \pi \hbar)^2} \sum_{\sigma} \int_0^{\infty} d p \,2 p f_{\sigma}(p)
  = \frac{1}{(2 \pi \hbar)^2} \left[\mu^2 + \frac{(\pi T)^2}{3}\right] \label{eq:rel1}
\end{equation}
with $f_{\sigma}(x) = [e^{(x - \sigma \mu)/T} + 1]^{-1}$ for particles ($\sigma = +1$) and antiparticles ($\sigma = -1$).

In high energy physics, the local charge current~\eqref{eq:semiclassical5} is employed to analyze the CVE.
Indeed, from Eq.~\eqref{eq:semiclassical5} we correctly reproduce the well-known fact
that $S = (1/3) \sigma_{\rm v}$ originates from the SVC, while $M = (2/3) \sigma_{\rm v}$ from the magnetic moment~\cite{PhysRevLett.113.182302}.
It is also consistent with the nonzero temperature part of the axial magnetic effect,
which is reciprocal to the CVE in the relativistic case, namely, the energy current is induced by an axial magnetic field~\cite{PhysRevD.88.071501,Buividovich_2015}.

In contrast, the transport charge current~\eqref{eq:semiclassical9} vanishes since $S = C = (1/3) \sigma_{\rm v}$.
In other words, the chiral vortical current is just the magnetization charge current, which cannot be measured in transport experiments.
This is consistent with the aforementioned fact that the distribution function of ${\tilde \epsilon}_\sigma({\bm p}, {\bm x})$ is the equilibrium one, because of the rotation symmetry.
The same is true even when the chiral imbalance is dynamically generated,
since the transport charge current vanishes regardless of the presence or absence of $\mu_5$.
It is a sharp contrast to the CME, where the charge current is proportional to $\mu_5$, leading to the NMR~\cite{NIELSEN1983389,PhysRevB.88.104412}.
We remark here that the CVE in condensed matter systems with the pseudo-Lorentz symmetry comes to the same conclusion
even if there exist multiple nodes with different energies.

\section{(Anisotropic) CVE in nonrelativistic systems} \label{sec:nonrel}
In nonrelativistic systems, a variety of spin-orbit couplings (SOCs) are allowed dependently on the crystal symmetry.
From the viewpoint of symmetry, among the $21$ noncentrosymmetric point groups, the possible candidates to yield the (anisotropic) CVE are the $11$ chiral point groups; $O$, $T$, $D_n$, $C_n (n = 2, 3, 4, 6)$, and $C_1$.
However, the CVE can never take place if the contributions from the SVC and Berry curvature cancel each other.
In this section, we analyze the (anisotropic) CVE for the above $11$ point groups.

In the point groups $O$ and $T$, only the isotropic CVE is allowed but cannot be observed experimentally in the low-energy regime.
We consider a spinful one-band Hamiltonian ${\hc H}({\bm p}) = g_0({\bm p}) + {\bm g}({\bm p}) \cdot {\bm \sigma}$.
The eigenvalues are $\epsilon_{\sigma}({\bm p}) = g_0({\bm p}) + \sigma g({\bm p})$ [$\sigma = \pm1$], and the corresponding eigenstates are
$| u_+({\bm p}) \rangle = [\cos \Theta/2, e^{i \Phi} \sin \Theta/2]^{\rm T}$ and $| u_-({\bm p}) \rangle = [-e^{-i \Phi} \sin \Theta/2, \cos \Theta/2]^{\rm T}$,
where $g$, $\Theta$, and $\Phi$ are the polar coordinates of ${\bm g}({\bm p})$.
We obtain ${\bm s}_{\sigma}({\bm p}) = \sigma (\hbar/2) {\bm g}({\bm p})/g({\bm p})$,
and ${\bm \Omega}_{\sigma}({\bm p}) = -\sigma \hbar^2 \sin \Theta {\bm \nabla}_p \Theta \times {\bm \nabla}_p \Phi/2$.
The symmetry allows the forms of $g_0({\bm p}) = p^2/2 m$ and ${\bm g}({\bm p}) = \alpha {\bm p}$.
The only scalar parts of $S$ and $C$ in Eq.~\eqref{eq:semiclassical6} are nonzero and read
\begin{equation}
  S = C
  = \hbar \sgn \alpha \sum_{\sigma}\sigma  \int \frac{d^3 p}{(2 \pi \hbar)^3} f(\epsilon_{\sigma}({\bm p}))/3 p, \label{eq:nonrel1}
\end{equation}
where $f(\epsilon) = [e^{(\epsilon - \mu)/T} + 1]^{-1}$ is the Fermi distribution function.
Thus, the transport charge current vanishes since $S = C$ as for relativistic chiral fermions.

We comment on Ref.~\cite{PhysRevResearch.2.032021}.
The authors also showed that in {\it arbitrary} chiral point groups the transport charge current vanishes.
The underlying mechanism in their argument is, however, $S = C = 0$;
the former is because the SVC was {\it not} taken into account, and the latter comes from the assumption of the parabolic dispersion relations.
This assumption is not always valid, and in fact $C$ does not vanish.
As shown here, the cancellation of the contributions from the SVC and Berry curvature, $S = C \not= 0$, is essential for the vanishing transport charge current of the CVE.
Indeed, in the following we show that {\it some} chiral point groups can admit the nonzero transport charge current of the anisotropic CVE.

In the point groups $D_n$ and their subgroups $C_n$ and $C_1$, the anisotropic CVE can be observed experimentally.
As a representative, we consider $n$-type tellurium that belongs to the point group $D_3$.
Trigonal tellurium lacks the inversion symmetry and has attracted renewed interest in the context of the spin and orbital Edelstein effects~\cite{Yoda2015,Furukawa2017}
and Weyl semimetals~\cite{PhysRevLett.110.176401,PhysRevLett.114.206401,PhysRevB.95.125204,Ideue25530,PhysRevLett.124.136404}.
The bottom of the conduction bands are close to the $H$ and $H^{\prime}$ points
and are described by $g_0({\bm p}) = (p_x^2 + p_y^2)/2 m_1 + p_z^2/2 m_2$ and ${\bm g}({\bm p}) = [v_1 p_x, v_1 p_y, v_2 p_z]^{\rm T}$.
The material parameters were experimentally determined to be $v_1 = 3.42 \times 10^4~{\rm m/s}$, $v_2 = 1.08 \times 10^5~{\rm m/s}$~\cite{JPSJ.35.525},
$m_1 = 0.104 m_0$, and $m_2 = 0.0697 m_0$, where $m_0$ is the electron mass.
The signs of $v_1$ and $v_2$ are determined by the spin texture obtained by first-principles calculation~\cite{PhysRevLett.114.206401}.
Although it was pointed out in Ref.~\cite{PhysRevLett.114.206401} that the antisymmetric SOC is isotropic,
the obtained spin texture is tilted to the $p_z$ axis, which indicates $v_1 < v_2$.

\begin{figure}
  \centering
  \includegraphics[clip,width=0.48\textwidth]{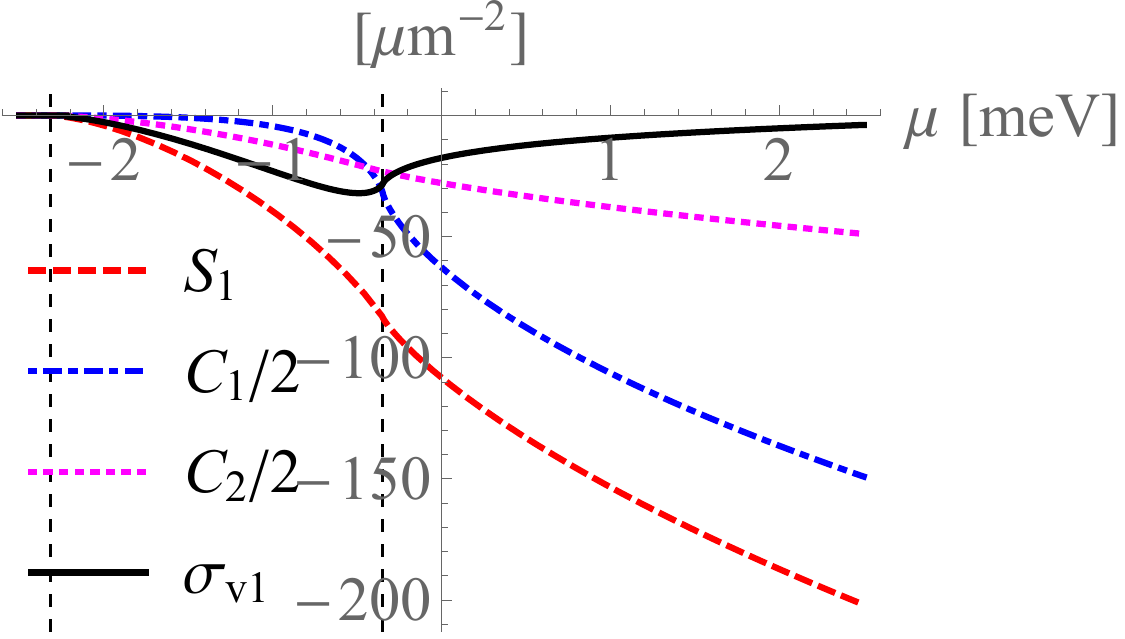}
  \caption{%
  In-plane chiral vortical conductivity $\sigma_{{\rm v} 1} = S_1 - (C_1 + C_2)/2$ (black solid line) as a function of the chemical potential $\mu$ for the effective model of $n$-type tellurium.
  The red dashed, blue dot-dashed, and magenta dotted lines represent $S_1$, $C_1/2$, and $C_2/2$, respectively.
  The black dashed lines correspond to the bottom of the conduction bands in the $p_x p_y$ plane, $-\epsilon_1 = -0.347~{\rm meV}$,
  and that in the $p_z$ axis, $-\epsilon_2 = -2.32~{\rm meV}$.%
  } \label{fig:cve}
\end{figure}

The coefficients $S$ and $C$ in Eq.~\eqref{eq:semiclassical6} are diagonal.
In particular, at zero temperature, they are written as
\begin{subequations} \begin{align}
  S_1
  = & {S^x}_x
  = {S^y}_y
  = -\frac{\sgn v_2}{\pi^2 \hbar^2 v_1 v_2} \int_{-1}^1 d x (1 + x^2) \notag \\
  & \times [E(x)]^2 \sqrt{1 + \mu/E(x)}, \label{eq:nonrel2a} \\
  S_2
  = & {S^z}_z
  = -\frac{2 \sgn v_2}{\pi^2 \hbar^2 v_1^2} \int_{-1}^1 d x (1 - x^2) \notag \\
  & \times [E(x)]^2 \sqrt{1 + \mu/E(x)}, \label{eq:nonrel2b} \\
  C_1
  = & C_{xx}
  = C_{yy}
  = S_2, \label{eq:nonrel2c} \\
  C_2
  = & C_{zz}
  = -\frac{4 \sgn v_2}{\pi^2 \hbar^2 v_2^2} \int_{-1}^1 d x x^2 \notag \\
  & \times [E(x)]^2 \sqrt{1 + \mu/E(x)}, \label{eq:nonrel2d}
\end{align} \label{eq:nonrel2}\end{subequations}
where we introduce $E(x) = [(1 - x^2)/\epsilon_1 + x^2/\epsilon_2]^{-1}$ with $\epsilon_1 = m_1 v_1^2/2 = 0.347~{\rm meV}$ and $\epsilon_2 = m_2 v_2^2/2 = 2.32~{\rm meV}$.
We have also taken into account the valley degrees of freedom, namely, the $H$ and $H^{\prime}$ points.
The symmetry allows the anisotropic CVE characterized by
\begin{subequations} 
\begin{align}
  j^{x (y)}
  = & q \sigma_{{\rm v} 1} \omega^{x (y)}, \label{eq:nonrel3a} \\
  j^z 
  = & q \sigma_{{\rm v} 2} \omega^z. \label{eq:nonrel3b}
\end{align} \label{eq:nonrel3}\end{subequations}
In Fig.~\ref{fig:cve}, we show the in-plane component $\sigma_{{\rm v} 1} = S_1 - (C_1 + C_2)/2$ as a function of the chemical potential.
Since the rotation symmetry is absent in the $x (y)$ axis, the in-plane vorticity drives the system out of equilibrium.
Indeed, the left-hand side of the Boltzmann equation~\eqref{eq:semiclassical2} is nonzero.
The in-plane component $\sigma_{{\rm v} 1}$ attains the negatively maximum value $\sigma_{{\rm v} 1 {\rm max}} = -32~{\rm \mu m}^{-2}$
near the bottom of the conduction bands in the $p_x p_y$ plane.
On the other hand, the out-of-plane component $\sigma_{{\rm v} 2} = S_2 - C_1$ always vanish.
This is because the system is in equilibrium when the vorticity is in the $z$ direction.
Note that, even if a vanishing transport charge current cannot be measured in transport experiments,
the local one is still physical and can be measured with local probes such as a superconducting quantum interference device.

\section{Discussion and summary} \label{sec:discussion}
We discuss the experimental setup to observe the anisotropic CVE obtained in the previous section.
In order to realize the situation where a rotating fluid is observed in an inertial frame, we use a surface acoustic wave,
which is rotational deformation localized at a surface.
It was theoretically proposed~\cite{PhysRevB.87.180402} and recently experimentally observed~\cite{PhysRevLett.119.077202}
that the spin current can be induced by the surface acoustic wave via the SVC.
The induced vorticity is expressed as $\omega^x(t, y, z) = (k_0^2 u_0/2 c_{\rm t}) e^{-k_{\rm t} z + i (k y - k_0 t)}$ with $k_0 = \xi c_{\rm t} k$, $k_{\rm t} = \sqrt{1 - \xi^2} k$, where $u_0$ is the amplitude of a mechanical resonator,
and $\xi$ is a parameter determined by the transverse sound velocity $c_{\rm t}$ and longitudinal sound velocity $c_{\rm l}$~\cite{9780750626330}.
The localized ac charge current with the frequency $k_0$ flows in the $x$ direction.
We apply the low frequency about $k_0/2 \pi = 100~{\rm MHz}$ to ensure that the vorticity is almost static and uniform.
With typical parameters, $u_0 = 1~{\rm nm}, c_{\rm t} = 2000~{\rm m/s}$~\cite{PhysRevB.87.180402},
we estimate the amplitude to be $\omega^x = k_0^2 u_0/2 c_{\rm t} = 1 \times 10^5~{\rm s}^{-1}$
and the maximum charge current to be $j_{\rm max}^x = q \sigma_{{\rm v} 1 {\rm max}} \omega^x = 0.5~{\rm A}/{\rm m}^2$.

We can also observe the anisotropic CVE simply by rotating a tellurium crystal with the angular velocity ${\bm \omega} = (\omega^x,0,0)$.
In this case, it is useful to adopt the corotating frame with the material.
Under the transformation from the inertial frame to the rotating frame, physical quantities are in general observed as different values.
The charge current parallel to the vorticity (or the angular velocity of the rotating frame), however, is irrelevant to choice of the frame;
e.g., the chiral vortical current of chiral fermions is unchanged even in the frame corotating with the fluid~\cite{PhysRevD.99.085014}.
Hence, the in-plane CVE in Eq.~\eqref{eq:nonrel3a} can be observed when the contacts rotate together with the material.

To summarize, we have formulated the CVE and GVE using the semiclassical theory of wave packet dynamics.
We have taken the SVC into account and calculated the transport charge current
by subtracting the magnetization one from the Noether local one~\cite{PhysRevResearch.2.032021,PhysRevLett.97.026603,PhysRevB.55.2344}.
In a relativistic system of chiral fermions, the transport charge current always vanishes since the contributions from the SVC and the Berry curvature cancel each other.
This result implies that the CVE cannot be observed in transport experiments
even when the chiral imbalance is generated applying parallel electric and magnetic fields~\cite{NIELSEN1983389,PhysRevB.88.104412}.
The conclusion is the same in condensed matter systems with the pseudo-Lorentz symmetry.
We have also demonstrated that the anisotropic CVE can be observed in nonrelativistic systems
that belong to the point groups $D_n$, $C_n (n = 2, 3, 4, 6)$, and $C_1$, such as $n$-type tellurium.
The charge current induced by a surface acoustic wave is estimated to be of the order of $0.1~{\rm A}/{\rm m}^2$.
%--- Acknowledgments
\begin{acknowledgments}
  We thank R.~Toshio for discussing the GVE and sharing their manuscript~\cite{PhysRevResearch.2.032021}.
  We also appreciate fruitful discussion with M.~Hirayama on tellurium and with H.~Watanabe on antisymmetric SOCs.
  This work was supported by the Japan Society for the Promotion of Science KAKENHI (Grant No.~JP18K13508).
\end{acknowledgments}
%--- Appendices
\appendix
\begin{widetext}
\section{Anisotropic CVE in $n$-type tellurium} \label{app:nonrel}
For the effective model of $n$-type tellurium, the nonzero components of $S$ and $C$ in Eq.~\eqref{eq:semiclassical6} are
\begin{subequations} \begin{align}
  S_1
  = & {S^x}_x
  = {S^y}_y
  = 2 \times \sum_{\sigma} \int \frac{d^3 p}{(2 \pi \hbar)^3} \frac{\sigma \hbar v_1 (v_1^2 p_y^2 + v_2^2 p_z^2)}{2 [g({\bm p})]^3} f(\epsilon_{\sigma}({\bm p})) \notag \\
  = & \frac{\sgn v_2}{8 \pi^2 \hbar^2 v_1 v_2} \int_0^{\pi} \sin \theta d \theta (\sin^2 \theta + 2 \cos^2 \theta)
  \sum_{\sigma}\sigma  \int_0^{\infty} d \epsilon \epsilon f(\epsilon^2/4 E(\theta) + \sigma \epsilon), \label{eq:nonrel4a} \\
  S_2
  = & {S^z}_z
  = 2 \times \sum_{\sigma} \int \frac{d^3 p}{(2 \pi \hbar)^3} \frac{ \sigma\hbar v_1^2 v_2 (p_x^2 + p_y^2)}{2 [g({\bm p})]^3} f(\epsilon_{\sigma}({\bm p}))
  = \frac{\sgn v_2}{4 \pi^2 \hbar^2 v_1^2} \int_0^{\pi} \sin \theta d \theta \sin^2 \theta
  \sum_{\sigma} \sigma \int_0^{\infty} d \epsilon \epsilon f(\epsilon^2/4 E(\theta) + \sigma \epsilon), \label{eq:nonrel4b} \\
  C_1
  = & C_{xx}
  = C_{yy}
  = 2 \times \sum_{\sigma} \int \frac{d^3 p}{(2 \pi \hbar)^3} \frac{\sigma \hbar v_1^2 v_2 p_x^2}{[g({\bm p})]^3} f(\epsilon_{\sigma}({\bm p}))
  = \frac{\sgn v_2}{4 \pi^2 \hbar^2 v_1^2} \int_0^{\pi} \sin \theta d \theta \sin^2 \theta
  \sum_{\sigma}\sigma  \int_0^{\infty} d \epsilon \epsilon f(\epsilon^2/4 E(\theta) + \sigma \epsilon)
  = S_2, \label{eq:nonrel4c} \\
  C_2
  = & C_{zz}
  = 2 \times \sum_{\sigma} \int \frac{d^3 p}{(2 \pi \hbar)^3} \frac{\sigma \hbar v_1^2 v_2 p_z^2}{[g({\bm p})]^3} f(\epsilon_{\sigma}({\bm p}))
  = \frac{\sgn v_2}{2 \pi^2 \hbar^2 v_2^2} \int_0^{\pi} \sin \theta d \theta \cos^2 \theta
  \sum_{\sigma}\sigma  \int_0^{\infty} d \epsilon \epsilon f(\epsilon^2/4 E(\theta) + \sigma \epsilon). \label{eq:nonrel4d}
\end{align} \label{eq:nonrel4}\end{subequations}
Here, the factor $2$ originates from the valley degrees of freedom, namely, the $H$ and $H^{\prime}$ points.
We have introduced new variables as $|v_1| p_x = \epsilon \sin \theta \cos \phi$, $|v_1| p_y = \epsilon \sin \theta \sin \phi$, and $|v_2| p_z = \epsilon \cos \theta$.
At zero temperature, using
\begin{equation}
  \sum_{\sigma} \sigma \int_0^{\infty} d \epsilon \epsilon f(\epsilon^2/4 E(\theta) + \sigma \epsilon)
  = -8 [E(\theta)]^2 \sqrt{1 + \mu/E(\theta)}, \label{eq:nonrel5}
\end{equation}
we obtain Eq.~\eqref{eq:nonrel2}.
\end{widetext}
%--- References
%Control: production of eprint (1) enabled
%
\end{document}